\documentclass[10pt,a4paper]{article}

\usepackage{amssymb,amsmath}
\usepackage{graphicx}
\usepackage{natbib}
\usepackage{url}
\usepackage{algorithm}
\usepackage{algpseudocode}
\algnewcommand{\LineComment}[1]{\State \(\triangleright\) #1}

\algnewcommand{\Parameter}[1]{\State #1}

\algnewcommand\algorithmicinput{\textbf{INPUT:}}
\algnewcommand\INPUT{\item[\algorithmicinput]}
\algnewcommand\algorithmicoutput{\textbf{OUTPUT:}}
\algnewcommand\OUTPUT{\item[\algorithmicoutput]}
\algnewcommand\algorithmicglobalvariables{\textbf{GLOBAL VARIABLES:}}
\algnewcommand\GLOBALVARIABLES{\item[\algorithmicglobalvariables]}

\algnewcommand\And{\textbf{and}}
\algdef{SE}[DOWHILE]{Do}{doWhile}{\algorithmicdo}[1]{\algorithmicwhile\ #1}%

\makeatletter
\newcommand{\StatexIndent}[1][3]{%
  \setlength\@tempdima{\algorithmicindent}%
  \Statex\hskip\dimexpr#1\@tempdima\relax}
\algdef{S}[WHILE]{WhileNoDo}[1]{\algorithmicwhile\ #1}%
\makeatother

\newcommand{\TOOLNAME}{$E^2FM$}

\begin{document}

\title{\TOOLNAME: an encrypted and compressed full-text index for collections of 
genomic sequences\thanks{This work was published on \textit{Bioinformatics}, 
doi: 10.1093/bioinformatics/btx313}}
\author{Ferdinando Montecuollo\\
CRESSI, Universit\`{a} ``Luigi Vanvitelli'', Napoli, 80133 Italy
\and
Giovannni Schmid\\
ICAR, Consiglio Nazionale delle Ricerche, Napoli, 80131, Italy
\and
Roberto Tagliaferri\\
DISA-MIS, Universit\`{a} di Salerno, Fisciano, 84084,Italy}

\date{}

\maketitle

\begin{abstract}
Next Generation Sequencing (NGS) platforms and, more generally, high-throughput 
technologies are giving rise to an exponential growth in the size of nucleotide 
sequence databases. Moreover, many emerging applications of nucleotide datasets 
-- as those related to personalized medicine -- require the compliance with 
regulations about the storage and processing of sensitive data. \\
We have designed and carefully engineered \TOOLNAME-index, a new full-text index 
in minute space which was optimized for compressing and encrypting nucleotide 
sequence collections in FASTA format and for performing fast pattern-search queries. 
\TOOLNAME-index allows to build self-indexes which occupy till to 1/20 of the storage
required by the input FASTA file, thus permitting to save about 95\% of storage when 
indexing collections of highly similar sequences; moreover, it can exactly search the 
built indexes for patterns in times ranging from few milliseconds to a few hundreds 
milliseconds, depending on pattern length.\\
Supplementary material and supporting datasets are available through Bioinformatics 
Online and https://figshare.com/s/6246ee9c1bd730a8bf6e.
\end{abstract}

\section{Introduction}
Next Generation Sequencing (NGS) platforms and, more generally, high-throu\-ghput 
technologies are giving rise to an exponential growth in the size of nucleotide
sequence databases. Moreover, many emerging applications of nucleotide datasets 
-- as those related to personalized medicine -- require the compliance with 
regulations about the storage and processing of sensitive data.  

Great efforts have been made in the last years to obtain compressed representations 
of genomic sequences. 
Referential genome compression algorithms \citep{saha2016nrgc} can compress very 
efficiently a large set of similar sequences by aligning each element in the set 
onto the reference sequence and by encoding mismatches between them.
However this is a compression strategy inapplicable to experiments for which a 
reference sequence is not clearly defined (\textit{metagenomics}) or entirely absent 
(\textit{de-novo discovery})\citep{yanovsky2011recoil}.\\
Reference-free compression does not suffer the above limitations, and various  
algorithms have been introduced that obtain excellent results in terms of compression 
ratio, search efficiency or sequence alignment.  
Methods based on the {\em Burrows Wheeler Transform} (BWT) \citep{burrows1994block} 
are particularly interesting in this respect because they support the construction 
of special {\em indices} (e.g. ausiliary data on the compressed text) that permit 
substring queries directly on compressed text, so avoiding the overhead in both 
space and time due to the decompression of data.\\ 
{\em Bzip2} \cite{bzip2} is a compressor that gets very good compression 
ratios on most files thanks to the BWT followed by a {\em Move-To-Front} (MTF) 
transform \citep{ryabko1980data} and {\em Huffman Coding} (HC) \citep{cormen2009introduction}. 
Bzip2 is suitable for compressing single files, not multiple files (i.e. file archives).
This is because it divides a text into blocks of size between 100 and 900 kbytes 
and then compresses each block separately. This way the BWT, which basically acts 
as a preprocessor for the compressors MTF and HC, is only able to take advantage 
of local similarities in the data.  \\
Indeed, the BWT operates a permutation on the input text which results in grouping 
its symbols into substrings of like letters. In \cite{mantaci2005extension} it is
shown that extending the BWT to a collection of sequences allows a much better 
space-efficiency than the technique used in Bzip2, because of redundancy arising 
from long-range correlations in the data.
On the other hand, running the BWT on large datasets is memory and CPU consuming.

In \cite{bauer2011lightweight}  fast and RAM-efficient methods capable of 
computing the BWT of sequence collections of the size encountered in human whole 
genome sequencing experiments are described. This approach is implemented in {\em BEETL} 
(Burrows-Wheeler Extended Tool Library) \citep{cox2012large}, a suite of applications 
for building and manipulating the BWT of collections of DNA sequences. 
Using BEETL the redundancy present in large-scale genomic sequence datasets can be 
fully exploited by generic second-stage compressors such as Bzip2. 
If compared to the naive use of Bzip2, this results in more than a four-fold increase 
in compression efficiency. 
However BEETL does not offer any data indexing, thus performing pattern-search queries
on datasets requires their decompression.

{\em Bowtie} \citep{langmead2009ultrafast} is a memory-efficient tool for aligning 
short DNA sequence reads to large genomes. For the human genome, Bowtie can align 
more than 25 million ``short reads'' (35 base-pair) per CPU-hour with a memory 
footprint of only about 1.3 gigabytes (GB), which allows to run Bowtie on a computer 
with 2 GB of RAM. Bowtie builds a self-index of a (single) reference sequence, and alignes the 
DNA sequence reads with respect to such index. It employs a {\em Full-text Minute-space} 
(FM-) index \citep{ferragina2000opportunistic}, a data structure based on the BWT which 
allows compression of the input text while still permitting fast substring queries. 

In this article we present \TOOLNAME-index (Extended and Encrypted Full-text 
Minute space index), a tool designed for storing in compressed and encrypted form
massive collections of genomic sequences and performing fast pattern-search queries 
on them. 
Our approach is similar to Bowtie, in that it makes use of a FM-index. However,
our goal was to get efficiently an {\em encrypted} self-index for a {\em whole 
collection} of genomic sequences, rather than aligning the collection items to a 
single indexed reference sequence.
At least at our knowledge, a natively encrypted self-index approach has been presented 
neither in the data and text mining literature nor, more specifically, for genome analysis. 
A traditional way to get confidentiality protection for compressed data is the so 
called ``compress-then-encrypt'' paradigm, in which encryption is performed  through 
a dedicated algorithm after data compression steps have taken place. 
For example, compress-then-encrypt methods have been documented in the ZIP File 
Format Specification since version 5.2 \cite{2013pavlov}, and an AES-based 
standard has been developed for WinZip \cite{winzip} and is used also in 
other file archivers [e.g. 7-Zip \cite{2013pavlov}].\\
However, the ``compress-then-encrypt'' approach has the drawback that one must first 
decrypt the file or the archive in its entirety before to operate on the compressed 
file.
For massive data amounts, as in case of nucleotide datasets, this can lead to big 
downgrades in performance; moreover, it exposes data on disk during operations, 
which can be an issue if the databases are in outsourcing or in multi-tenants environments 
(e.g. cloud environments). \\ 
A more interesting approach stems from new generation filesystems with built-in
encryption, like ZFS \citep{bonwick2003zettabyte}. 
Using such a filesystem one could put a cleartext index on disk, getting it automatically 
encrypted and decrypted at the filesystem level. However, this 
approach requires the reorganization of collections of genomic sequences as one 
or more filesystems. Moreover, with this approach each successful authentication results 
in data being transparently unencrypted for the user.    
Conversely, an encrypted index entails a two-layered data protection because of:
(i) the password required to access the system or database where the index is 
stored and (ii) the key required to decrypt the index.

The paper is organized as follows. Section \ref{sec:systemandmethods} gives an 
overview of the main features of \TOOLNAME, alongside with the computational methods
which make possible such features. Section \ref{sec:algorithms} illustrates more
deeply some core algorithms implemented in \TOOLNAME, in order to point out some 
important differences of our approach with respect to current computing techniques 
for genome analysis. 
Section \ref{sec:results} reports some of the numerical experiments we ran to assess 
the performance of our tool versus a state-of-the-art index.
Section \ref{sec:security} discusses the resiliency of our encryption method with
respect to some prominent data breach attacks in the context of genomic dataset
services. 
Finally, Section \ref{sec:conclusion} sums up the main features of \TOOLNAME and
sketches out future works.

\section{System and Methods}
\label{sec:systemandmethods}
The \TOOLNAME-index is an open-source C++ tool that makes it practical to compute an 
encrypted self-index of large collections of genomic sequences in FASTA format. 
This way genomic datasets can be stored in both encrypted and compressed form on disk.
Nonetheless pattern-searching such as ``count'' and ``locate'' queries can be performed 
efficiently on these datasets, and such queries require the decryption in main memory 
of the portion of data which is effectively involved in the query.  

The \TOOLNAME-index achieves compression thanks to a pipeline of BWT, MTF and RLE0 
transformations like an FM-index. Unlike an FM-index, however, \TOOLNAME perfoms its
computations on an ``extended and scrambled'' alphabet. As we are going to detail 
in the following, this can results in a better compression and, moreover, can offer 
some confidentiality protection to data during their processing in main memory.

One other main feature of \TOOLNAME  is that it natively implements an 
efficient encryption method based on the \textit{Salsa20} stream cipher \cite{salsa20}. 
Since the cipher operates separately on each single block of the index,
only the blocks of the index which are affected by a pattern-search query are 
decrypted at run time in memory and searched in compressed form.

Our C++ encryption routines interface with the Salsa20 assembly code available on 
the estream portfolio \cite{estream}. 
This, alongside with the use  of vector instructions included in modern CPUs 
and multithreaded programming strategies, allows to speed up cryptographic operations 
and minimize both index construction and pattern search times.

\TOOLNAME  has a simple command line interface that allows also non-experienced 
users to easily perform basic operations such as the generation of an encryption 
key, the construction of an index and the execution of pattern searching queries. 
It is also possibile to extract subsequences of collection items by supplying in 
input the desired item index and the start position and length of the required 
subsequence.    

\subsection{Encoding collections of genomic sequences}
Both genomic sequences and search patterns are strings in the standard ISO/IU\-PAC 
nucleid acid notation. The IUPAC alphabet $\Sigma_{IUPAC}$
contains the five symbols $\{A,C,G,T,U\}$, corresponding to DNA and RNA bases, 
plus a set of 12 additional symbols representing possible ambiguities caused by 
sequencing machines errors or inaccuracy (for example, the ``B''symbol stands 
for ``not A'').

Given a collection of genomic sequences $C=\{S_1,S_2,\cdots,S_n\}$, it is often 
the case that only a subset of the symbols in $\Sigma_{IUPAC}$ are actually required
to encode $C$. 
Thus, the first operation performed by \TOOLNAME consists in constructing $\Sigma$,
where  $\Sigma$ is the alphabet containing 
the only symbols in 
$\bigcup_{i=1}^{n} S_i$ plus the two more special symbols $\$$ and $\&$.

Let us now consider the \textit{$k$-extension alphabet} $\Sigma^k$ of $\Sigma$, that 
is the k-fold Cartesian product $\Sigma^k = \Sigma \times \dots \Sigma$. The symbols 
of $\Sigma^k$ are $k$-length strings of $\Sigma$ symbols and are called \textit{$k$-mers}.
Finally, let $s^k$ denote the $k$-mer obtained  by repeating $k$ times the same 
symbol $s$.

The whole collection $C$ can be represented as a single string of symbols in 
$\Sigma^k$ as follows:

\begin{figure*}[!htbp]

\includegraphics[width=\textwidth]{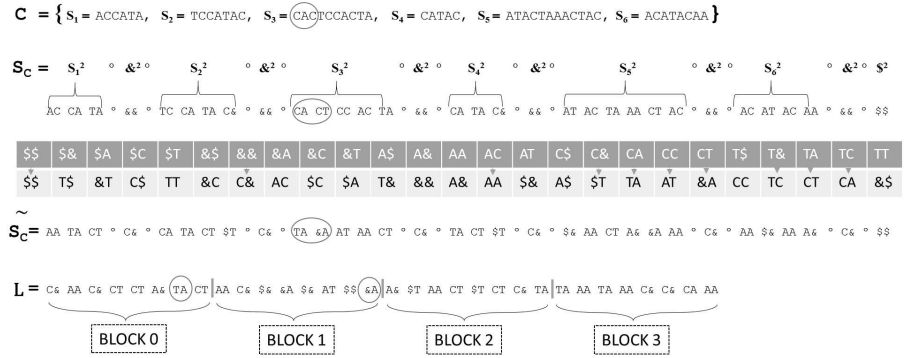}
\caption{\small{An illustrative example of the index construction process is 
depicted here top-down from the input $C$ to the output $L$, as follows. 
The collection $C$ of six genomic sequences and its extended sequence $S_C$ with 
respect to $\Sigma^2$ (i.e. for $k=2$). 
A two rows table representing the pseudo-random permutation performed on $\Sigma^2$, 
and the resulting scrambled sequence $\widetilde{S_C}$. 
The BWT $L$ of $\widetilde{S_C}$, and its partitioning in blocks of size $bs=8$.}}

\label{fig:CodingExample}
\end{figure*}
\begin{itemize}
\item Each collection item $S_i$ is coded as a sequence of $\Sigma^k$ symbols 
($k$-mers), by taking its characters in blocks of $k$. If the length of $S_i$ is 
not a multiple of $k$, then $S_i$ is first right padded with as many $\&$ symbols 
as required; 
\item The so obtained $S_i^k$ are concatenated in the string $S_C$ of $k$-mers
given by: 
$$S_C = S_1^k \circ \&^k \circ S_2^k \circ \&^k \circ \cdots \circ \&^k \circ 
    S_n^k\circ \&^k \circ \$^k$$
\end{itemize}
As we stated previously, \TOOLNAME actually computes the BWT with respect 
to a ``scrambled'' alphabet $\widetilde{\Sigma^k}$. 
A cryptographically secure pseudo-random generator based on \textit{Salsa20} is 
used to change the order of $\Sigma^k$ symbols.
Then the scrambled extended sequence $\widetilde{S_C}$ of $S_C$ is computed by
re-encoding the symbols in $S_C$ with respect to their new ordering in 
$\widetilde{\Sigma^k}$.

As we are going to show through some numerical experiments in Section \ref{sec:results},
suitable values for the parameter $k$ results both in better compression ratios 
and faster BWT computation times. Computation is faster because the number of 
elements in $\widetilde{S_C}$ is about $1/k$ times the total number of bases 
in $C$. On the other hand, compression ratios are improved since alphabet extension
results in a MTF compression closer to \textit{universal coding} performance
\citep{bentley1986locally}, i.e. that of a compressor having a compression 
ratio which differs at most for a constant factor from that of the optimal prefix 
code\footnote{In our case this happens only up to small values of $k$ (tipically
$k=4$ or $k=5$), since increasing $k$ results also in more metadata composing
the index.}. 
Moreover, scrambling the extended alphabet offers some confidentiality protection
during data processing in main memory, as we are going to detail in Section
\ref{sec:security}. \\

Figure \ref{fig:CodingExample} illustrates the overall process of the ``scrambled''
BWT computation for a small collection $C$ of short genomic sequences, an extension 
factor $k=2$ and a block size $bs=8$. This simple example serves also to show how
extending and scrambling the alphabet $\Sigma$ results in the splitting and scattering 
of subsequences of nucleotide basis in the original genomic dataset.
For example, the \texttt{CAC} subsequence (which could be seen as a \textit{codon} 
codifying for an amino acid) at the beginning of $S_3$ sequence in $C$ is splitted 
into the two 2-mers \texttt{CA} and \texttt{CT} in $S_C$. In turn, these two strings 
are replaced respectively by \texttt{TA} and \texttt{\&A} in $\widetilde{S_C}$  
because of the scrambling operation. Finally, they are moved in two different 
blocks of the index thanks to the BWT.

\subsection{Computing the BWT}
The previous encoding technique allows to compute the BWT on the entire given 
collection of genomic sequences, thus exploiting ``runs'' of like letters not only
in the same sequence but overall in the collection.
In order to minimize computing time and memory footprint we designed and implemented
a new multi-threading algorithm using a ``block-based'' approach similar to that introduced 
in \cite{karkkainen2007fast}, where a suitable set of ``splitters'' is  chosen so that the 
ranges of suffixes delimited by them can be ordered separately.
Unlike \cite{karkkainen2007fast}, we choose the splitters by observing the statistical 
properties exhibited by $k$-mers in the input genome. 
Our approach stems from this simple observation: given a genomic sequence $s$ in 
$\Sigma$, its rotations can be evenly distributed over contiguous ranges of $\Sigma^k$ 
symbols, on the basis of the lexicographic order of their first $k$ characters.

In order to obtain an additional performance increment, we reserved a special 
treatment to \textit{long repetitions} of the same character, which make the ordering 
very difficult. The above-mentioned ranges containing long repetitions are split 
into several subranges, which in turn are separately ordered (see Section \label{sec:algorithms}
for further details).
An example of such repetitions are the very long patterns of $N$ (aNy) symbols 
often occurring in genomic reference sequences due to reading errors.
Our strategy results in a significant speedup in BWT calculation of genomic 
sequences already on systems with only a couple of quad-core CPUs.

\subsection{Constructing the encrypted index}
\label{subsection:constructing_encrypted_index}
Given $L=BWT(\widetilde{S_C})$, $L$ is first splitted in a sequence of fixed-size 
blocks and superblocks, as it happens in the original FM-index. The block size 
is provided as the input parameter $bs$, whilst the superblocks size is computed 
from $bs$ so that each superblock contains exactly 16 blocks.
Afterwards, the $L$ symbols falling in each block are remapped with respect to the 
smallest alphabet required for that particular block.
Lastly, the sequence of symbols in each block is encoded as follows:
\begin{itemize}
\item a Move To Front transform (MTF) is followed by a Run Length Encoding of zeros 
(RLE0);
\item a {\em keystream} from the secret key $k_{enc}$ and the block number is
computed thanks to a pseudorandom number generator based on the \textit{Salsa20} 
stream cipher \cite{salsa20});
\item the output of the MTF-RLE0 pipeline is encrypted with a XOR-style cipher, 
using the keystream computed at the previous step;
\item the encrypted data is coded by using the minimum number of bits needed to 
represent the alphabet of each block.
\end{itemize}
Notice that we do not use the Multiple Tables Huffman (MTH) encoding as in the 
FM-index, because of the large memory footprint of its related decoding tables.

\subsection{Searching for patterns} 
\label{subsection:searching_for_patterns}
Once the \TOOLNAME-index on a collection $C$ has been constructed, it can be used 
to find the occurrences of a pattern $P$ within the items of $C$.  
Like the original FM-index, \TOOLNAME implements an exact pattern search through 
the backward search algorithm given in \cite{ferragina2000opportunistic}. However, 
we re-engineered that algorithm in order to obtain a good performance on the extended 
alphabet.
Compared to the extension $\Sigma^k$, a search for a single pattern $P \in \Sigma$
is indeed equivalent to search for a set of super-patterns. This set consists of
super-patterns being associated with each of the $k$ possible displacements
($d= 0, 1, \dots, k-1$) between $P$ and the symbols of $\Sigma^k$.
Table \ref{tab:SuperPatterns} illustrates this circumstance for $
\Sigma=\{ \texttt{\$, \&, A, C, G, N, T }\}$, $k=4$ and $P=\texttt{ACGAACTGA}$.
Symbol \texttt{?} denotes any single character of $\Sigma$, with the only
constraint that the special symbols \texttt{\$} and \texttt{\&} cannot occur in 
a super-pattern.
It is easy to see that the set related to each displacement is composed by exactly
$(|\Sigma|-2)^{(k - |P| \mod k)}$ elements. Thus, the total
number of super-patterns that must be searched in order to look for $P$ is given
by
\begin{equation}
\label{eq:num_spatterns}
k (|\Sigma|-2)^{(k - |P| \mod k)} \ ,
\end{equation}
which can be a significant value for some choices of $\Sigma$, $k$, and $|P|$.
For example, in the case illustrated in Table \ref{tab:SuperPatterns}, the set
of super-patterns corresponding to $d=0$ is composed of the 125 strings of 12
characters having the required pattern as prefix. Thus, looking for $P=\texttt{ACGAACTGA}$
in a naive way would correspond to search for a total number of 500 super-patterns.

\begin{table}[htbp]
\centering
\caption{Example of super-patterns with variable symbols.}
\label{tab:SuperPatterns}
\begin{minipage}{0.98 \textwidth}
 \begin{center}
    \begin{tabular}{|r||r|r|r|}
    \hline
    Displacement     &   \multicolumn{3}{c|}{Super-patterns}  \\
    \hline
     0 & \texttt{ACGA}     & \texttt{ACTG} & \texttt{A???}\\
     1 & \texttt{?ACG}     & \texttt{AACT} & \texttt{GA??}\\
     2 & \texttt{??AC}     & \texttt{GAAC} & \texttt{TGA?}\\
     3 & \texttt{???A}     & \texttt{CGAA} & \texttt{CTGA}\\
     \hline
    \end{tabular}
\end{center}

\footnotetext{Searching for pattern $\texttt{ACGAACTGA}$ in the alphabet
$\Sigma =$ $\{ \texttt{\$, \&, A, C, G, N, T }\}$ corresponds to search for the 
above set of super-patterns in $\Sigma^4$. Symbol \texttt{?} denotes any single 
character of $\Sigma$.}
\end{minipage}
\end{table}
On the other hand, performing the backward search algorithm a considerable number 
of times involves a large number of block readings from disk, which in turn can 
significantly degrade performance. In order to avoid this problem we designed and
implemented a backward search algorithm which is optimized for super-patterns with 
variable super-characters, like those shown in Table \ref{tab:SuperPatterns}. 
We are going to describe in depth this algorithm in the next section.

\section{Algorithms}
\label{sec:algorithms}
As told in the introduction, the two main differences of \TOOLNAME with respect to a 
standard FM-index are that: (i) it is designed to operate on entire collections of 
genomic sequences and, (ii) it has built-in an advanced encryption mechanism.   
However, \TOOLNAME makes also use of different optimization strategies in both
its design and implementation for improving its space and time efficiencies.
In order to better understand these last differences, we thoroughly describe some 
core algorithms of \TOOLNAME in the following subsections.

\subsection{Index construction}
\begin{figure}[htbp]
\begin{center}
\includegraphics[scale=0.25]{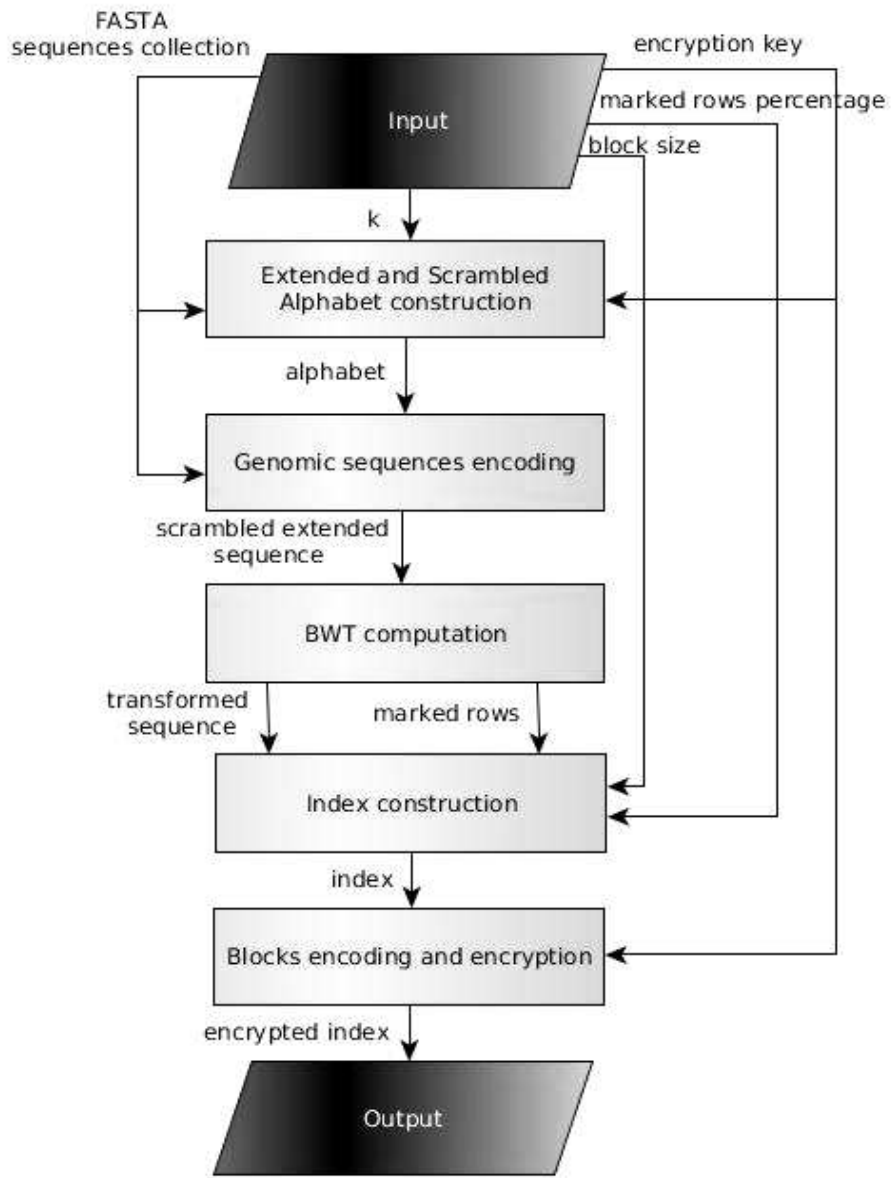}
\caption{\small{An overall view of the \TOOLNAME-index building process}}
\label{fig:indexBuildOverallView}
\end{center}
\end{figure}

An overall sketch of the index building process is given in Fig. \ref{fig:indexBuildOverallView}.

\noindent The \TOOLNAME-index is constructed taking in input the following five 
parameters:
\begin{itemize}
\item the full path of a collection of sequences in FASTA format;
\item an integer $k$, indicating the desired {\em extension order} of $\Sigma$, 
which determines the $k$-fold Cartesian product $\Sigma^k = \Sigma \times \dots 
\Sigma$;
\item the block size $bs$, that is the fixed number of $BWT$ symbols that fall in 
each data block;
\item the percentage of marked rows (as in the FM-index, marked rows allow for 
``locate'' queries);
\item an enciphering/deciphering key $encryption key$ consisting of a 64 byte array,
for a total size of 512 bits.
\end{itemize}

\subsubsection{Algorithms for scrambling the extended alphabet}
As we stated previously, only a subset of the symbols in $\Sigma_{IUPAC}$ is
actually required to encode $C$. Therefore, first of all Algorithm 
\ref{alg:ScrambledAlphabetConstruction} builds the alphabet $\Sigma$, containing 
the only $\Sigma_{IUPAC}$ symbols actually in $C$, plus the two special symbols 
$\$$ and $\&$. 
Then Algorithm \ref{alg:ScrambledAlphabetConstruction} computes a scrambling key, 
which is a permutation of the $\Sigma^k$ elements that defines their ordering in 
$\widetilde{\Sigma^k}$. The permutation is  computed with the {\em Fisher-Yates 
shuffle} \citep{durstenfeld1964algorithm}, using a pseudo-random number generator 
based on the Salsa20 cypher which is initialised with the first 32 bytes of the 
\TOOLNAME-index encryption key.

 
\begin{algorithm} 
\caption{Construction of the ``extended and scrambled'' alphabet}
\label{alg:ScrambledAlphabetConstruction}
\begin{algorithmic}[1]

\small{
\Function{ScrambledAlphabetConstruction}{$C$,$k$,$k_{enc}$}
\LineComment{Retrieve $\Sigma_{IUPAC}$ symbols actually present in C items}
\State $collectionSymbols \gets retrieveSymbols(C)$  \Comment{multi-threaded}
\LineComment{Build $\Sigma$ as union of collection symbols and special symbols}
\State $\Sigma \gets collectionSymbols \cup \{\texttt{\$}\} \cup \{\texttt{\&}\}$
\LineComment{Compute the extended alphabet's cardinality}
\State $eac \gets |\Sigma|^k $;
\LineComment{Initialize scrambling key}
\For{$i \gets 0$ To $eac-1$}
      \State $sk[i] \gets i$;
\EndFor
\LineComment{Initialize a pseudo-casual number generator based on Salsa20 cypher}

\State $salsa20Key \gets k_{enc}[0:31]$; \Comment{first 32 bytes of $k_{enc}$}
\State $salsa20Nonce \gets 0$; \Comment{nonce is always equal to 0}
\State $rnd \gets$ new $RandomGenerator(salsa20Key,salsa20Nonce)$; 
\LineComment{Shuffle the $sk$ elements by the Fisher-Yates algorithm (Knuth shuffle),}
\LineComment{excluding the first one.}
\For{$i \gets eac$ DownTo $1$}
      \Do
	\State $toSwapWith \gets rnd.nextInt(i)$;
      \doWhile{$toSwapWith = 0$};
      \LineComment{Swap element in place $i-1$ with that in place $toSwapWith$}
      \State $tmp \gets sk[i-1]$;
      \State $sk[i-1] \gets sk[toSwapWith]$;
      \State $sk[toSwapWith] \gets tmp$;       
\EndFor    
\State \Return new $ScrambledAlphabet(\Sigma,k,sk)$;  

\EndFunction

}
\end{algorithmic}
\end{algorithm}

\subsubsection{Algorithms for the BWT computation}
The main algorithm (see Algorithm \ref{alg:BWTComputation}) takes in input the 
sequence $\widetilde{S_C}$, the scrambled extended alphabet $\widetilde{\Sigma^k}$, 
the number $nt$ of sorting threads and the number $nr$ of ranges of $\widetilde{\Sigma^k}$. 
These ranges are a set of intervals of contiguous 
$\widetilde{\Sigma^k}$ characters that constitute a partition of the alphabet.
Since BWT computation requires the ordering of $\widetilde{S_C}$ rotations, Algorithm 
\ref{alg:BWTComputation} first partitions $\widetilde{\Sigma^k}$ into $nr$ ranges 
of contiguous characters and then distributes such rotations among the aforecited 
ranges.
Finally, it distributes those ranges among the $nt$ sorting threads through a {\em 
greedy} algorithm \citep{cormen2009introduction} named $split$, in order to balance 
the workload. 
Sorting in each range is performed through the \textit{multi-key quick sort} 
algorithm \citep{bentley1997fast}. 
Finally, the $computeBWT$ algorithm merges the $nt$ sorting results obtained in
the previous step and computes the BWT. For further details please refer to the 
\texttt{FastBWTransformer} C++ class source code.

\begin{algorithm}
\caption{BWT computation}\label{alg:BWTComputation}
\begin{algorithmic}[1]

\small{
\Function{BWTComputation}{$\widetilde{S_C}$,$\widetilde{\Sigma^k}$,$nt$,$nr$}

\LineComment{Fill the array of ranges (single-thread step)}
\LineComment{Ranges are right-open interval)}
\State $rangesWidth=|\widetilde{\Sigma^k}|/nr$;
\State $i=0$;
\WhileNoDo{$i < |\widetilde{\Sigma^k}|$}      
    \State $R[i].firstCharacter \gets i$;            
    \State $R[i].lastCharacter \gets i+rangesWidth$;    
    \State $R[i].rotations= \emptyset$;
    \State $i = i + rangesWidth$;
\EndWhile;

\LineComment{Distribute rotations among ranges (multi-thread step)}
\State $distributeRotations(\widetilde{S_C},R,nt)$;

\LineComment{Distribute ranges containing at least one rotation among the $nt$ threads,}
\LineComment{splitting the array $R$ in $nt$ subarrays (single-thread step)}
\LineComment{divided by $nt-1$ splitters}
\State $splitters=split(R,nt)$;

\LineComment{Sort rotations in each range (multi-thread step)}
\State $sort(R,nt,splitters)$;

\LineComment{Compute BWT, merging sort results}
\State $result=computeBWT(R)$;

\State \Return result;

\EndFunction

}
\end{algorithmic}
\end{algorithm}

\subsubsection{Algorithms for block encoding and encryption}
As detailed in subsection \ref{subsection:constructing_encrypted_index}, the $BWT$ 
returned by Algorithm \ref{alg:BWTComputation} is splitted in blocks of size $bs$.    
Then Algorithm \ref{alg:BlocksTextEncoding} implements the second encryption step: 
it applies a XOR-style cypher to data of each block, using the keystream produced 
by a pseudorandom number generator based on the \textit{Salsa20} stream cipher 
\cite{salsa20}. The pseudorandom generator is initialised with the last 32 bytes 
of the $k_{enc}$ (the first 32 were used for scrambling) and a \textit{nonce} (number 
used only once) corresponding to the specific block number. A different nonce is 
used for different blocks, in order to realize a non-deterministic encryption and 
thwart chosen plaintext attacks (see Section \ref{sec:security}).
After the encryption, algorithm \ref{alg:BlocksTextEncoding} encodes the block 
symbols using the smallest number of bits capable to represent them.
\begin{algorithm}
\caption{Blocks text encoding}\label{alg:BlocksTextEncoding}
\begin{algorithmic}[1]

\small{
\Function{EncodeBlockText}{$blockNumber$,$blockText$,$blockLength$,$k_{enc}$}
\LineComment{\textbf{Generate keystream}}
\LineComment{Initialize a pseudo-casual number generator based on Salsa20 cypher}
\State $salsa20Key \gets k_{enc}[32:63]$; \Comment{last 32 bytes of $k_{enc}$}
\State $salsa20Nonce \gets blockNumber$; \Comment{nonce is equal to blockNumber}
\State $rnd \gets$ new $RandomGenerator(salsa20Key,salsa20Nonce)$;
\LineComment{Generate a number less than block's alphabet size for each item of the blockText array.}
\LineComment{Block's alphabet contains only symbols actually occurring within the block.}
\For{$i \gets 0$ To $blockLength-1$}      
	\State $keyStream[i] \gets rnd.nextInt(blockAlphaSize)$;            
\EndFor
\LineComment{\textbf{Encode block's text}}
\LineComment{Apply to block's text the MTF transformation and RLE0 encoding }
\LineComment{After RLE0 block size has been reduced to compressedLength}
\State $transformedText \gets RLE0(MTF(blockText))$;
\State $compressedLength \gets length(transformedText)$;
\For{$i \gets 0$ To $compressedLength-1$}      
	\State $keyStream[i] \gets rnd.nextInt(blockAlphaSize)$;            
\EndFor
\LineComment{Allocate result vector, whose size is equal to $compressedLength$}
\State $result \gets new uint32_t[compressedLength]$;
\LineComment{Encrypt the RLE0's result using the previously generated keystream}
\For{$i \gets 0$ To $compressedLength-1$}      
        \LineComment(\% is modulus operator)
	\State $result[i] \gets (transformedText[i]+keystream[i])\%blockaphaSize)$;            
\EndFor
\State \Return result;  
\EndFunction

}
\end{algorithmic}
\end{algorithm}

Algorithm \ref{alg:BlocksTextEncoding} is actually a simplified version of that 
implemented in \TOOLNAME, which has been optimized to perform the MTF, RLE0 and 
encryption tasks at once on each block of text.
For further details, please refer to the \texttt{Bucket} C++ class source code.

\subsection{Algorithms for pattern search}
The overall pattern search strategy is summed up in Algorithm \ref{alg:SuperPatternSearch}.
As we previously said, pattern search takes place in the extended and scrambled 
string $L$ resulting from the BWT computation. Thus, the first operation consists 
in computing the super-patterns corresponding to the required pattern with respect 
to the scrambled alphabet $\widetilde{\Sigma^k}$. This work is performed by the 
\texttt{computeSuperPat\-terns} function. It produces exactly $k$ super-patterns,
one for each possibile displacement between the required pattern and the indexed 
data (see Table \ref{tab:SuperPatterns}).

As it should be clear from Table \ref{tab:SuperPatterns}, variable super-characters 
can occur just in the first and/or last position of a super-pattern. Actually, 
a variable super-character in the first position can be managed through one more 
iteration of the backward search algorithm. Indeed such super-character is matched 
by $backwardSearch$ against the super-characters that are compatible with its own 
mask\footnote{For example, consider the super-pattern $?ACG - ACCT - GA??$ in 
Table \ref{tab:SuperPatterns} and suppose that $backwardSearch$ until $ACCT$ 
returned the range of three rotations having the following last super-characters: 
$TCAA$, $CACG$, $CATT$. These super-characters are BWT elements and they precede 
$ACCT$ in the indexed string. In this case $backwardSearch$ returns the only 
rotation which is compatible with the mask $?ACG$, that is the rotation corresponding 
to $CACG$.}.

Thus, it remains to describe the inner working of our algorithm for a super-pattern 
with only the last super-character of variable type.
Let $P=P_0P_1...P_{m-1}$ $= \hat{P} P_{m-1}$ be a super-pattern with $P_i \in \Sigma^k$,
and where $P_{m-1}$ is its unique variable symbol.
Searching for $\hat{P}$ requires a single execution of the backward search algorithm,
and results in the range of rows with consecutive indexes $[\hat{sp}, \hat{ep}]$
in the array of suffixes \citep{ferragina2000opportunistic}. On the other hand, it
is easy to show that the rows in $[\hat{sp}, \hat{ep}]$ having $P_{m-1}$ in their
position $m-1$ are all and only the suffixes having as prefix the pattern $P$. Thus,
an efficient way to find $P$ consists in checking if the character in position
$m-1$ for each of the rows $[\hat{sp}, \hat{ep}]$ is encompassed in the variable
symbol $P_{m-1}$. Such check is performed by function {\em CheckLastChar} (see
Algorithm \ref{alg:CheckLastChar}), which uses the standard algorithms Locate
and Extract of the FM-index \citep{ferragina2000opportunistic} and returns the
position of the entire pattern $P$ if such pattern exists, the null string otherwise.
The $displacement$ function returns the super-pattern displacement, as described 
in subsection \ref{subsection:searching_for_patterns} and shown in table 
\ref{tab:SuperPatterns}.

We have optimized the above algorithms through a multi-threading strategy, so that 
the backward search of different super-patterns can be distributed among multiple 
threads. For details, please refer to the \texttt{EFMCollection} 
and the \texttt{EFMIndex} C++ class source code.

\begin{algorithm}
\caption{\small{SuperPatternSearch: an optimized algorithm to search for patterns 
over a k-extension alphabet $\Sigma^k$.}}
\label{alg:SuperPatternSearch}
\begin{algorithmic}[1]
\small{
\Function{SS-search}{$originalPattern$}
\State positions=[]; \Comment{Positions of the pattern occurrences}
\State superPatterns=computeSuperPatterns($originalPattern$);
\For{P in superPatterns}
    \State $m \gets length(P)$;
    \State $\hat{P}=P_0P_1P_{m-2}$;
    \State $\tilde{P} \gets P_{m-1}$;
    \State $[\hat{sp},\hat{ep}] \gets backwardSearch(\hat{P})$;
    \For{i in $[\hat{sp},\hat{ep}]$}
        \State pos=CheckLastChar($i$,$\tilde{P}$,$m$);
	\If{ pos is not null }
 	  \State $d \gets  displacement(P)$; \Comment{Displacement of the super-pattern}
	  \State add(positions,pos*k+d);\Comment{k is the alphabet extension order}
        \EndIf
    \EndFor
\EndFor
\State \textbf{return} positions;
\EndFunction
}
\end{algorithmic}
\end{algorithm}

\begin{algorithm}
\caption{\small{{\em CheckLastChar}: a function called by algorithm {\em SuperPatternSearch}
in order to verify if a row $i$ satisfying $\hat{P}$ also satisfies $P$}}
\label{alg:CheckLastChar}
\begin{algorithmic}[1]
\small{
\Function{CheckLastChar}{$i$,$P$,$m$}
\State $pos \gets Locate(i)$;
\State $c \gets Extract(pos + m-1)$;
\If{c \textbf{like} $P_{m-1}$}
   \State \textbf{return} $pos$; \Comment{The position pos is also that of the entire pattern P}
\Else
   \State \textbf{return} null; \Comment{no match}
\EndIf
\EndFunction
}
\end{algorithmic}
\end{algorithm}

%
%
%

\section{Results}
\label{sec:results}
We ran a comprehensive set of functional and numerical tests in order to 
assess the reliability and the performance of our prototypal C++ implementation 
of \TOOLNAME. For the experimental setup we proceeded as follows.

First of all, we  built the
\textit{consensus} sequences\footnote{These are sequences of nucleotides in FASTA 
format obtained by applying to a chromosome reference sequence the DNA variations 
appearing in a specific individual.} related to chromosomes 1, 11 and 20 of 50 
individuals from the {\em 1000 Genomes Project}.
In order to build each consensus sequence we first downloaded the corresponding 
individual alignment data in BAM format from the {\em 1000 Genomes Project} FTP 
site, and then we supplied it in input to the \texttt{mpileup} tool of the
\textit{Samtools} suite \cite{samtools}. We used this kind of sequences to measure
the time required to construct the index, its compression ratio and the time spent
in searching for patterns of different lengths.\\

Secondly, we built \textit{pseudo-random} sequences related to chromosomes 11, 20,
and to a portion of 500 Kbases of chromosome 20, respectively for 100 and 500
individuals. This kind of sequences were obtained by applying single mutations, 
insertions and deletions to the corresponding chromosome reference sequence in the 
human genome bank HS37D5, a variant of the GRCh37 human genome assembly used by the 
{\em 1000 Genomes Project}.   
For this purpose we have built a tool which pseudo-randomly selects (with uniform 
distribution) mutations, insertions and deletions in a way that the \textit{mutation 
rate} is equal to $0.1\%$, the \textit{in-del rate} is equal to $0.013\%$ and the 
\textit{in-del length} varies in the interval $[1-16]$. According to \cite{mullaney2010small}
these are indeed the genetic changes observed on average among different individuals
of the human species. We used the collections of 100 individuals to measure, as 
before, the time required to construct the index, its compression ratio and the 
time spent in searching for patterns. Instead, the collections
of 500 individuals were used to measure the \textit{speedup} of \TOOLNAME versus 
the number of threads, and the memory footprint (as percentage of loaded 
blocks) of our tool during pattern searches.  
 
Finally, we measured the performance of \TOOLNAME on the single entire human genome, 
by considering the collection of all the human chromosomes contained into HS37D5.
This was a sort of ``stress testing'', especially in case of compression ratios,
since our tool has been designed to exploit the similarities among collection items.
However, these results are of some significance if compared with those obtained 
using our \textit{reference tool} (see below).

For each of the three above set of tests, we compared the performance of our 
prototype with a reference tool obtained from a state-of-the-art library for 
creating self-indexes, namely the {\em Sdsl C++ library} \cite{sdsl}. 
This library implements some \textit{succinct data structures} \citep{jacobson1988succint}
that can be used to construct self-indexes like {\em Compressed Suffix Arrays} 
(CSA) and {\em wavelet tree} FM-indexes. We had to extend the wavelet tree FM-index 
supplied by such library in order to manage collections of items and to report 
sequence-relative locations. In doing that we used the same approach described in
Section \ref{sec:systemandmethods}, but with a separator consisting in the single 
special character ``$\#$''.  

We ran our tests on different computing platforms, in order to evaluate somehow 
also the influence of the operating environment (amount/type of physical resources, 
operating system, virtualization technologies, etc.). 

A first set of tests was run on a virtual machine hosted by a Red Hat Enterprise 
Virtualization 3.4 system with 196 GB of RAM and 4 Intel(R) Xeon(R) CPU E5-2697 
v2 @ 2.70GHz 6-core processors (RHEV34 for short). This machine had Intel(R) 
Hyperthreading(R) technology enabled. 

A second set of tests was run on a laptop with Ubuntu Desktop 16.04 LTS, 8GB of 
RAM and an Intel(R) Core(TM) i7-4500U dual-core CPU @ 1.80GHz (LAUD16 for short). 

A third and latest set of test was run on a Ubuntu Server 16.04 virtual machine 
hosted by a {\em cloud service} provider and configured with 140 GB of RAM and 2 
Intel Xeon(R) E5-2673 v3 @ 2.40GHz 10-core processors (CLUS16 for short). For this 
machine Intel(R) Hyperthreading(R) technology was not enabled.

The following sections summarize the main results. For coherence the results reported 
here and through the supplementary material published at {\em Bioinformatics online} 
are all related to the CLUS16 operating platform. 
Numerical results obtained on the RHEV34 computing platform are instead available through 
{\em Figshare}. 
Finally, on the LAUD16 platform we were able to run only a subset of tests because of
the limited amount of memory; these results are just discussed in the following
without any published supporting dataset.

\subsection{Index construction performance}
\begin{figure}
\begin{center}
\includegraphics[width=0.98\textwidth]{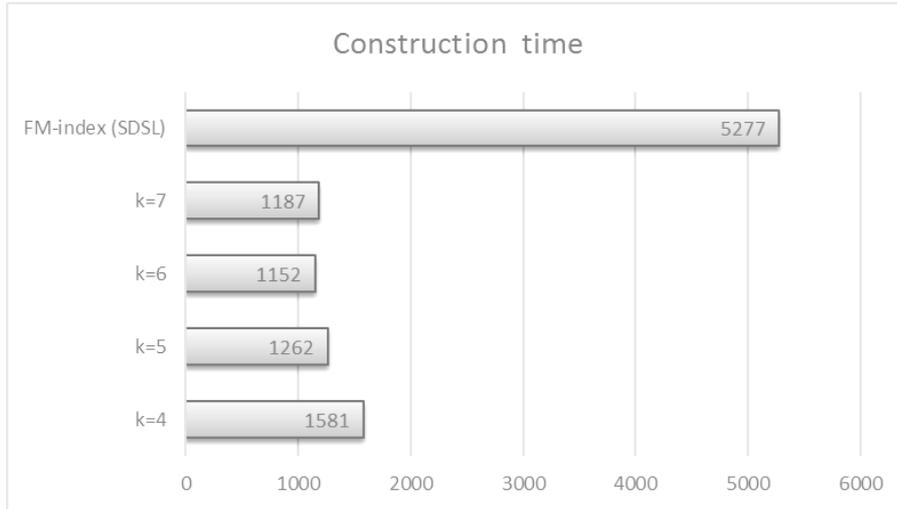}
\caption{\small{Comparison between the construction time (in seconds) of \TOOLNAME 
($k=4,5,6,7$) and that of the reference Sdsl FM-index tool on a human chromosomes 11 
collection of 50 consensus sequences.}}
\label{fig:collection_chr11_ct}
\end{center}
\end{figure}

All the tests measuring the time required for the construction of indices in main 
memory show that \TOOLNAME greatly outperforms the Sdsl FM-index, despite the fact that 
this last does not suffer the overhead due to encryption. This is because \TOOLNAME 
takes advantage of the multi-threaded Algorithm \ref{alg:BWTComputation} for the
BWT computation.
For example, as shown in Figure \ref{fig:collection_chr11_ct}, the indexing 
of the 6.28 GiB collection composed of 50 consensus sequences for the human 
chromosome 11 required less than 20 minutes with \TOOLNAME for $k=7$ and about 90 
minutes with the FM-index.\\
By comparing all the tests of this kind performed on the two different computing 
platforms  RHEV34 and CLUS16 we can conclude that building \TOOLNAME was about five 
time faster than building the reference tool in the best case ($k=6$, RHEV34) and more 
than three time faster in the worst case ($k=4$, CLUS16). Instead, on LAUD16 
\TOOLNAME run faster than the Sdsl FM-index already with two running threads,
meaning that our index has a good performance also on computing platforms with
limited resources.
Besides some influence by the computing environment, a key role here for the 
performance of \TOOLNAME is played by the alphabet's extension factor $k$. 
This is because the BWT computation is by far the most demanding computing task 
during the construction of \TOOLNAME, and the load of such computation increases 
as $k$ decreases. On the one hand, indeed, bigger $k$-mers have statistically less 
occurence in data, while on the other hand the complexity of a sorting problem 
(like the BWT) increases with the number of items to sort.      

In order to measure the effectivenes of our multithreading approach for Algorithm 
\ref{alg:BWTComputation} we run also some tests to measure the \textit{speedup}
of \TOOLNAME with respect to the number of running threads on the CLUS16 platform. 
They show (see Bioinformatics Online) that speedup scales significally until the 
gain resulting from splitting the sorting workload in subtasks is reduced by the 
costs due to synchronization.

\subsection{Compression ratios}
\begin{figure}
\begin{center}
\includegraphics[width=0.98\textwidth]{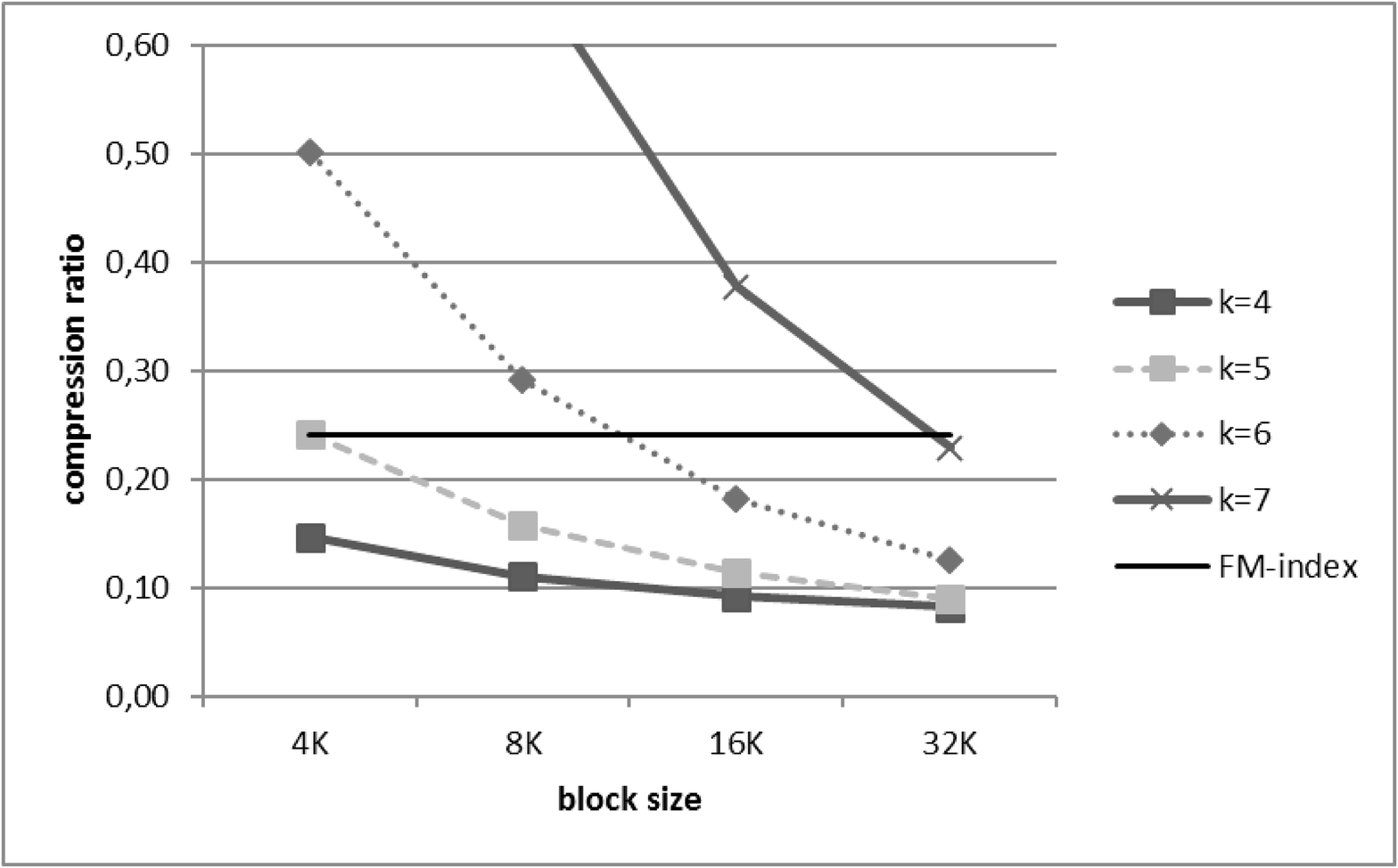}
\caption{\small{Compression ratios of \TOOLNAME versus the reference FM-index tool 
(SDSL) for a human chromosome 11 collection of 50 consensus sequences.}}
\label{fig:collection_chr11_cr}
\end{center}
\end{figure}

Figure \ref{fig:collection_chr11_cr} shows the compression ratios achieved with 
\TOOLNAME versus those got with the reference tool on the previous collection of 
50 consensus sequences for the human chromosome 11.
In this case the compression ratio of 24\% achieved with the Sdsl FM-index was more 
than halved by the best compression achieved with \TOOLNAME, which for $k=4$ and a 
block size $bs=32K$ resulted in a compression ratio less than 10\% . In this case the 
original 6.28 GiB data resulted in about 0.52 GiB of indexed and encrypted data.

Similar results, as documented by the supplementary material, were observed in all 
these kinds of tests and on all the tested computing platforms. 
Instead, \TOOLNAME slightly outperformed the reference tool on the whole human 
genome collection. However, this is a natural consequence of the fact that the 
redundancy between items in this collection is very low.

Overall, these results show that the compression ratios achieved with
\TOOLNAME\\ decrease with increasing block size values $bs$ and that, for a fixed $bs$, 
smaller values of $k$ result in better (i.e. smaller) compression ratios.


\subsection{Pattern search performance}
\label{sec:pattern_search_performance}
\begin{figure}
\begin{center}
\includegraphics[width=0.98\textwidth]{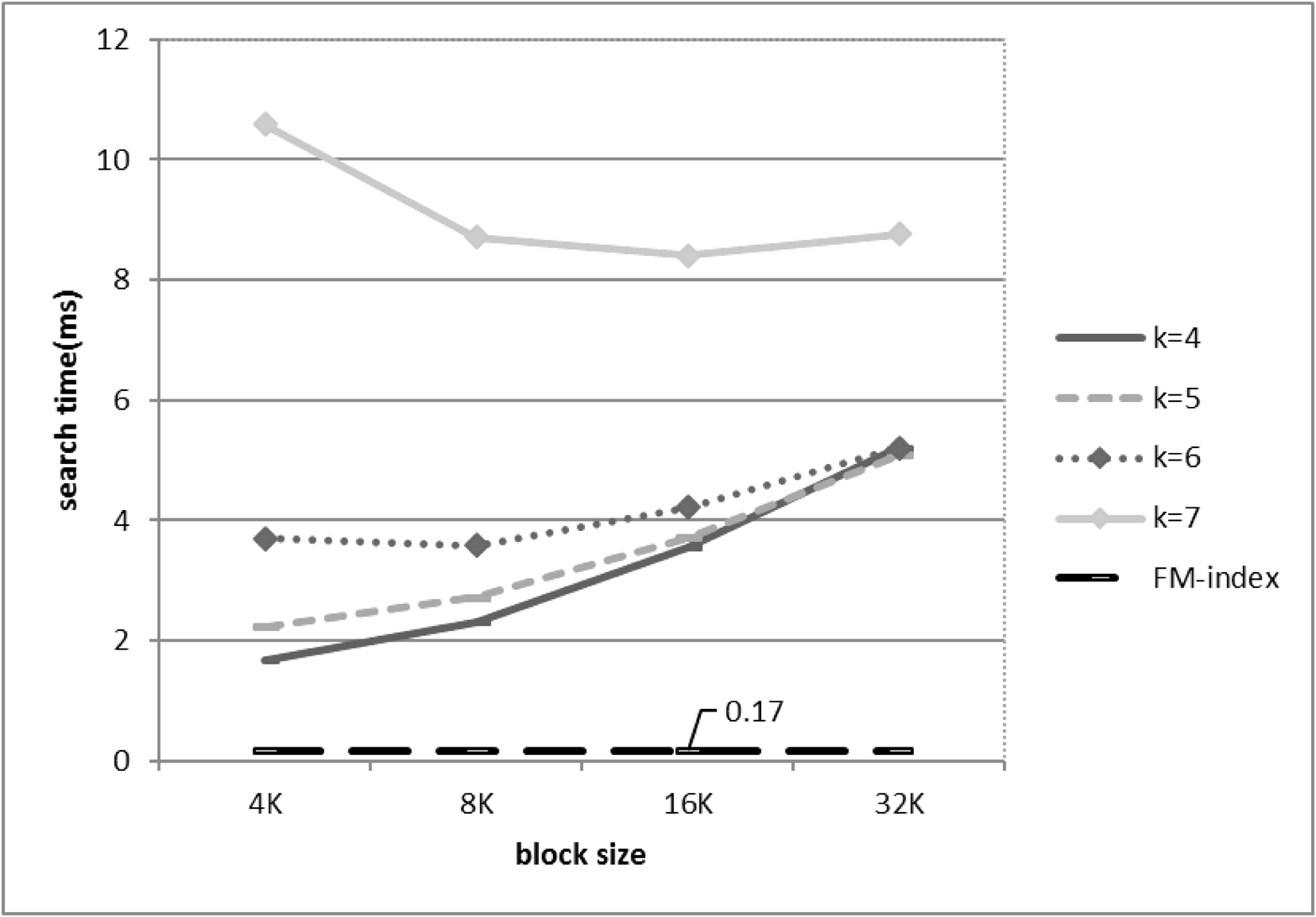}
\caption{\small{Mean searching times with respect to the pattern lengths of 20, 
50, 100, 200 and 500 bases for a human chromosome 11 collection of 50 
consensus sequences.}}
\label{fig:chr11_coll50_st}
\end{center}
\end{figure}

In this kind of tests we measured the pattern searching time of \TOOLNAME
versus that of the Sdsl FM-index.
The set of tests executed on RHEV34 measure the performance achieved in search 
operations by selecting at random 500 patterns of different lengths (15, 20, 
50, 100, 200 and 500 bases), and by computing -- for each set of patterns having 
the same length -- the median of the time spent to report the occurrence of each 
pattern in the set.\\
However, the length of patterns is usually unknown during a pattern search 
analysis; it will be rather one of the outcomes of the study. 
For this reason, in the subsequent set of tests performed on CLUS16 we decided
to measure the mean of the time spent in searching for patterns computed with 
respect to all the different pattern lengths.
Figure \ref{fig:chr11_coll50_st} shows the results for a collection of 100
pseudo-randomly chosen chromosomes 11 and a pattern of 50 basis. 

In almost all cases and on any platform \TOOLNAME performed better as $k$ 
decreased, but it was largerly outperformed by the reference tool. This is 
a clear consequence of the growing complexity of Algorithm \ref{alg:SuperPatternSearch}
with $k$. However, search times for retrieving each pattern occurrence were of the 
order of milliseconds in any case. Thus, this gap in performance for \TOOLNAME has 
no practical significance, except in case of very large sets of queries.

Pattern search performance depends on the block size in a more complex way.
As documented by the supplementary data, for short patterns searching times 
were roughly the same for all block sizes values, whilst a sensible change in time 
performance with varying block sizes was observed for medium-size and long patterns.
Overall, however, the tests show that nearly optimal searching times are achieved 
in all cases and on all platforms with a block size in the range $\{4K,16K\}$. 

An indirect measure of performance in pattern search is given by the number of 
blocks loaded in memory during this kind of tasks. Indeed, because of the lesser
I/O operations, performance improves as the number of loaded blocks decreases.
Thus, we carried out some tests on CLUS16, in order to measure the percentage of
loaded blocks for patterns of different lengths or different block sizes. 
These tests, as documented in the supplementary material at Bioinformatics Online, 
show that this percentage is very low. Actually \TOOLNAME was engineered to manage 
very efficiently data in memory. Indeed, on LAUD16 we were able to construct our 
index and perform pattern searching through it for genomic sequences of 1.5 GiB 
and more, whilst the Sdsl FM-index cannot be constructed. Similar behaviours were
observed also on the RHEV34 and CLUS16 platforms, as documented by the online 
supporting datasets, although of course for much bigger sequences. 

\section{Security considerations}
\label{sec:security}
Data breaches are becoming a major concern in information societies. 
The increasing relevance of digital processing and the diffusion of mobile and
outsourced computing are indeed weakening the role of traditional protection
mechanisms based on physical controls. Genomic databanks and related genome 
analysis services expose sensitive data and thus require adequate protection.
Many are the ongoing efforts to get more secure computing services, in particular
through advanced cryptographic protocols for performing on-line computations with
privacy protection for the users. For example, in \cite{shimizu2016efficient} {\em
additive homomorphic encryption} is used in order to conceal the sequence query 
and the region of interest when a user searches for information on a server
that stores a large indexed dictionary and employs the BWT for query operations.
However, as reported by some prominent risk analysis services (see for example
\url{www.breachlevelindex.com/}, Gemalto Data breach statistics), 
one main threat is nowdays represented by {\em data thefts}, and this is because
data is very often stored unencrypted on disk. \TOOLNAME has been designed and 
implemented for storing on disk large collections of genomic sequences in encrypted 
and compressed form. This way it can mitigate the risks subsequent to the theft 
of data; moreover, this protection is complementary to that offered by secure 
protocols for interacting with on-line databank services.  

As we have illustrated in the previous sections, \TOOLNAME natively implements a
very efficient encryption method based on the {\em Salsa20} stream cipher. 
As of 2017 there are no published attacks on Salsa20; moreover, the 15-round Salsa20 
was proven {\em 128-bit secure} against differential cryptanalysis \citep{mouha2013towards}.
We have also said (see Section \ref{sec:systemandmethods}) that \TOOLNAME offers 
some sort of confidentiality protection to data during their processing in main 
memory.
Since this feature can be useful in some kind of {\em multitenant} computing 
environments, like the cloud computing environments deployed by some providers,
we are going to sketch below some facts that corroborate our claim.  
   
It would be easy to show that the BWT computed on $\widetilde{S_C}$ results in a 
{\em poly-alphabetic substitution cipher} \citep{menezes2010handbook} that, for 
alphabets of suitable size and {\em homophonic} input data, can thwart exhaustive 
key-search attacks and cryptanalytic attacks based only on ciphertext knowledge 
({\em ciphertext-only attacks}) \citep{menezes2010handbook}.\\
The above argument seems not to apply to DNA sequences since they: (i) exhibit a 
strong structure, at least in some their parts, and (ii) provide large segments 
of available plaintext to a possible attacker (e.g. through the 1000Genome project).

As respect to (i) the simple example of Figure \ref{fig:CodingExample} suggests 
however that the ``extended and scrambled alphabet'' approach is able to break the 
strong regularities existing in some regions of DNA.
Because of the expansion in $k$-mers and the (unknown) reordering of $k$-mers
performed by the BWT, sequence of patterns in the plaintext are splitted in pieces
and these are scattered all over the index.\\
Actually, the {\em degree of homophony} $O$ in a plaintext $p$ can be measured by the 
number of possible choices for an ordered array of symbols of $p$ so to match the 
array of decreasing non-zero frequencies of occurence of symbols in $p$. 
That integer $O$ indeed represents the number of possible trials an attacker has 
to do in the worst case in order to find the right matching. 
We computed the degree of homophony for different values of $k$ in plaintexts 
$p$ given by the genomic data in input and expressed in symbols of the extended 
alphabet (i.e. the $k$-mers): it was of the order of $10^{22}$ already for $k=4$,
and it was orders of magnitude greater than  $10^{100}$ for $k \in \{5,6,7,8\}$.   

As respect to (ii), the key observation is that the attacker has to learn some 
{\em new} and {\em specific} genomic pattern or profile (e.g. a mutation in an 
individual), starting from the (first stage) ciphertext load in main memory during 
a pattern search and the knowledge deriving from publicly available genomic data.  
However, it has to face the following obstacles. The public available information
is only generic, and a specific pattern or profile can consist in a variation 
which could also affect large portions of the sequence (as it happens with any 
ins/del). Our tool was designed to work on {\em collections} of genomic sequence 
(e.g. a specific chromosome for a set on different individuals); because of the 
BWT way of processing such information will be scrambled and spread all over the
resulting (first stage) ciphertext. On average, the percentage of ciphertext loaded 
in memory is very low, as illustrated in subsection \ref{sec:pattern_search_performance},
so the attacker has to perform its statistical cryptanalysis on a lacking sample
of the ciphertext. Starting form the array of frequencies of symbols desumed by 
the knowledge deriving from publicly available genomic data, which is biased with
respect to the array of frequencies of the (unknown) plaintext, the attacker has 
to solve a combinatorial best matching problem with respect to the (poor) array of 
frequencies computed on the ciphertext loaded in memory. Finally, the attacker
has to choose the right set of symbols corresponding to the array of frequencies 
for the plaintext among the $O$ possible corrispondences due to the homophony in
the plaintext.

\section{Conclusion and future work}
\label{sec:conclusion}
\TOOLNAME-index is a new full-text index in minute space which was optimized for 
compressing and encrypting entire collections of genomic sequences and for performing 
fast pattern-search queries. 
\TOOLNAME has been developed in C++ using the vector instructions of modern CPUs 
and multithreaded programming strategies. Moreover, encryption routines interface 
with the assembly code of a state-of-the-art encryption tool, namely the Salsa20 
stream cipher. 
With \TOOLNAME  command line interface it is easy to perform  operations such as 
the generation of an encryption key, the construction of an index, the execution 
of pattern searching queries and the extraction of  subsequences of collection 
items. 

We ran a comprehensive test set to compare the performance of \TOOLNAME with a 
reference tool based on the FM-index. These tests show that \TOOLNAME takes much 
less time to be constructed and it greatly outperforms the reference tool in 
compression ratios. As respect to pattern search performance, \TOOLNAME  is reasonably 
worse than the reference tool, meaning that the implemented encryption mechanisms 
results in a very low overhead.
Besides, the heuristic following our experiments resulted in the following 
simple ``rule of thumb'' for the choice of the input parameters: 
\begin{itemize}
\item the greater is  $k \in \{4,5,6,7\}$ the better is the confidentiality 
protection for the data loaded in main memory during searching operation;   
\item choose the value $bs$ of the block size as follows: $bs=4K$ for maximum 
performance in pattern search operations, $bs=8K$ for a good performance, $bs=16K$
for a good compression and $bs=32K$ for maximum compression.
\end{itemize}  




We are working to the design of database management systems which extend and improve 
the features of \TOOLNAME. For example, we are studying key management algorithms 
for granting access to genomic data through a role-based access control policy. 
Other research concerns the extension of our pattern search algorithm to inexact 
sequence mapping, which is a main subject in bioinformatics.   


\bibliographystyle{natbib}
\bibliography{main.bib}

\end{document}